\documentclass[preprints,article,accept,moreauthors,pdftex]{Definitions/mdpi}

\firstpage{1}
\pubvolume{xx}
\issuenum{1}
\articlenumber{5}
\pubyear{2020}
\copyrightyear{2020}
\history{Received: date; Accepted: date; Published: date}

\Title{Search for double beta decay of $^{106}$Cd with an
enriched $^{106}$CdWO$_4$ crystal scintillator in coincidence with
CdWO$_4$ scintillation counters}

\Author {P.~Belli$^{1,2}$\orcidA{}, R.~Bernabei
$^{1,2}$*\orcidB{}, V.B.~Brudanin$^{3}$\orcidC{},
F.~Cappella$^{4,5}$\orcidD{}, V.~Caracciolo$^{1,2,6}$\orcidE{},
R.~Cerulli$^{1,2}$\orcidF{}, F.A.~Danevich$^{7}$\orcidG{},
A.~Incicchitti$^{4,5}$\orcidH{}, D.V.~Kasperovych$^{7}$\orcidI{},
V.R.~Klavdiienko$^{7}$\orcidJ{}, V.V.~Kobychev$^{7}$\orcidK{},
V. Merlo$^{1,2}$\orcidL{},O.G.~Polischuk$^{7}$\orcidM{},
V.I.~Tretyak$^{7}$\orcidN{} and
M.M.~Zarytskyy$^{7}$\orcidO{}}

\AuthorNames{P.~Belli, R.~Bernabei, V.B.~Brudanin, F.~Cappella, V.~Caracciolo, R.~Cerulli, F.A.~Danevich, A.~Incicchitti, D.V.~Kasperovych, V.R.~Klavdiienko, V.V.~Kobychev, V.~Merlo, O.G.~Polischuk, V.I.~Tretyak and M.M.~Zarytskyy}

\address{
$^{1}$ \quad INFN, sezione di Roma ``Tor Vergata'', I-00133 Rome, Italy\\

$^{2}$ \quad Dipartimento di Fisica, Universit\`{a} di Roma “Tor Vergata”, I-00133 Rome, Italy\\

$^{3}$ \quad Joint Institute for Nuclear Research, 141980 Dubna, Russia\\

$^{4}$ \quad INFN, sezione Roma “La Sapienza”, I-00185 Rome, Italy\\

$^{5}$ \quad Dipartimento di Fisica, Universit\`{a} di Roma ``La Sapienza'', I-00185 Rome, Italy \\

$^{6}$ \quad INFN, Laboratori Nazionali del Gran Sasso, 67100
Assergi (AQ), Italy \\

$^{7}$ \quad Institute for Nuclear Research of NASU, 03028 Kyiv, Ukraine
}

\corres{Correspondence: Dipartimento di Fisica, Universit\`{a} di Roma ``Tor Vergata'', I-00133 Rome, Italy.
E-mail address: rita.bernabei@roma2.infn.it (Rita Bernabei)}

\abstract{Studies on double beta decay processes in $^{106}$Cd
were performed by using a cadmium tungstate scintillator enriched
in $^{106}$Cd at 66\% ($^{106}$CdWO$_4$) with two CdWO$_4$
scintillation counters (with natural Cd composition). No effect
was observed in the data accumulated over 26033 h. New improved
half-life limits were set on the different channels and modes of
the $^{106}$Cd double beta decay at level of $\lim T_{1/2}\sim
10^{20}-10^{22}$ yr. The limit for the two neutrino electron
capture with positron emission in $^{106}$Cd to the ground state
of $^{106}$Pd, $T^{2\nu\mathrm{EC}\beta^+}_{1/2}\geq2.1\times
10^{21}$ yr, was set by the analysis of the $^{106}$CdWO$_4$ data
in coincidence with the energy release 511 keV in both CdWO$_4$
counters. The sensitivity approaches the theoretical predictions
for the decay half-life that are in the range
$T_{1/2}\sim10^{21}-10^{22}$ yr. The resonant neutrinoless
double-electron capture to the 2718 keV excited state of
$^{106}$Pd is restricted at the level of
$T^{0\nu\mathrm{2K}}_{1/2}\geq2.9\times10^{21}$ yr.}


\begin{document}

\section{Introduction}
\label{intro}

Observations of the neutrino oscillations suggest that the
neutrinos are massive, which calls for extension of the Standard
Model of particles and fields (SM). However, oscillation
experiments cannot determine the neutrino mass and the neutrino
mass hierarchy. One of the most promising tools to determine the
absolute neutrino mass scale and the neutrino mass hierarchy, the
nature of the neutrino (Dirac or Majorana particle?), to check the
lepton number conservation is double beta ($2\beta$) decay of
atomic nuclei, a process in which two electrons (or positrons) are
emitted simultaneously and nuclear charge changes by two units:
(A,Z)$\rightarrow$(A,Z$\pm 2$)
\cite{Giuliani:2012,Cremonesi:2014,Vergados:2016}. The
neutrinoless mode of the decay ($0\nu2\beta$) violates the lepton
number conservation law and it is possible if the neutrinos are
Majorana particles (particle is equal to its antiparticle). Being
a process beyond the SM, the $0\nu2\beta$ decay has the potential
to test the SM \cite{Bilenky:2015,DellOro:2016,Dolinski:2019}.
Moreover, the Majorana nature of the neutrino might shed light on
the Universe baryon asymmetry problem
\cite{Asaka:2005,Deppisch:2018}.

The two-neutrino $2\beta$ decay ($2\nu2\beta$) is a radioactive
process allowed in the SM with the longest half-lives ever
observed: $10^{18}$ -- $10^{24}$ yr. The $2\nu2\beta^-$ decay mode
has been detected in several nuclides \cite{Barabash:2020}. The
$0\nu2\beta$ decay is not observed. The most sensitive
$2\beta^-$-decay experiments quote half-life limits at level of
$T_{1/2} > (10^{24} - 10^{26}$) yr, which correspond to Majorana
neutrino mass limits in the range $\langle m_{\nu}\rangle < (0.1 -
0.7)$ eV. Probing the inverted hierarchy region of the neutrino
mass requires improved sensitivities of $2\beta^-$ experiments at
level of $\langle m_{\nu}\rangle \sim (0.02 - 0.05)$ eV (i.e.
half-life sensitivity in the range: $T_{1/2} \sim 10^{27}-10^{28}$
yr).

The sensitivity of the experiments in the search for ``double beta
plus'' processes: double electron capture (2EC), electron capture
with positron emission (EC$\beta^+$) and double positron decay
($2\beta^+$) is substantially lower, while the physical
lepton-number violating mechanisms of the neutrinoless 2EC,
EC$\beta^+$ and $2\beta^+$ processes are considered essentially the same as
for the decay with electrons emission. At the same time, there is
a motivation to search for the $0\nu$EC$\beta^+$ and
$0\nu2\beta^+$ decays owing to the potential to clarify the
possible contribution of the right-handed currents to the
$0\nu2\beta^-$ decay rate \cite{Hirsch:1994}, and an interesting
possibility of a resonant $0\nu$2EC process
\cite{Winter:1955a,Voloshin:1982,Bernabeu:1983,Krivoruchenko:2011}.

As for the allowed two-neutrino mode of the double beta plus
decay, there are claims of positive results (indication) for the
$2\nu$2EC radioactivity of three nuclides. The $2\nu$2EC decay of
$^{130}$Ba was claimed in two geochemical experiments where
anomaly in the isotopic concentrations of daughter xenon traces in
old barite (BaSO$_{4}$) minerals was interpreted as the sought
effect with the half-life $T_{1/2}=(2.16\pm0.52)\times 10^{21}$ yr
\cite{Meshik:2001}, and with $T_{1/2}=(6.0\pm 1.1)\times10^{20}$
yr in \cite{Pujol:2009}. In the analysis \cite{Meshik:2017} the
disagreement was explained by a possible cosmogenic contribution
with a conclusion that the result of \cite{Meshik:2001} is a more
reliable one. An indication on the $2\nu$2EC process in $^{78}$Kr
with the half-life $T_{1/2}=9.2^{+5.7}_{-2.9}\times10^{21}$ yr was
obtained with a proportional counter with a volume of 49 lt filled
by gas enriched in $^{78}$Kr to 99.81\% \cite{Gavrilyuk:2013}. The
value was then updated to $1.9^{+1.3}_{-0.8}\times10^{22}$ yr in
\cite{Ratkevich:2017}. Recently a detection of the $2\nu$2EC of
$^{124}$Xe with the half-life $(1.8 \pm 0.5) \times 10^{22}$ yr
was claimed in \cite{XENON:2019}. However, the indications of
$^{130}$Ba 2EC decay should be confirmed in direct counting
experiments, while the results for $^{78}$Kr and $^{124}$Xe need
to be confirmed with bigger statistics and very stable
experiments. Other allowed $2\nu$ decay channels with decrease of
the nuclear charge by two units, $2\nu$EC$\beta^+$ and
$2\nu2\beta^+$, are not observed yet.

The nuclide $^{106}$Cd is one of the most appealing candidates to
search for 2EC, EC$\beta^+$ and $2\beta^+$ decays with a long
history of studies (a review of the previous investigations reader
can find in Ref. \cite{Belli:2012}). The interest to $^{106}$Cd
can be explained by one of the biggest decay energy
$Q_{2\beta}=2775.39(10)$ keV \cite{Wang:2017}, comparatively high
isotopic abundance $\delta=1.245(22)$\% \cite{Meija:2016} and
possibility of gas centrifugation for enrichment, existing
technologies of cadmium purification, availability of
Cd-containing detectors to realize calorimetric experiments with a
high detection efficiency.

At present there are three running experiments searching for the
double beta decay of $^{106}$Cd: COBRA, TGV-2 and the present one.

The COBRA collaboration utilizes CdZnTe semiconductor detectors at
the Gran Sasso underground laboratory (Laboratori Nazionali del
Gran Sasso, LNGS). The experiment started with one
Cd$_{0.9}$Zn$_{0.1}$Te detector with mass of $\simeq $3 g, and one
CdTe detector ($\simeq $6 g) \cite{Kiel:2003}. CdZnTe detectors
are used in the current stage of the experiment
\cite{Ebert:2013,Ebert:2016}. The measurements resulted in the
half-life limits for several channels of $^{106}$Cd double beta
decay at level of $\sim10^{18}$ yr.

The main goal of the TGV-2 experiment, located at the Modane
underground laboratory, is search for $2\nu 2$EC decay of
$^{106}$Cd (a decay channel expected to be the fastest one) with
the help of 32 planar HPGe detectors with a total sensitive volume
$\approx400$ cm$^3$. In the first stage of the experiment, foils
of cadmium enriched in $^{106}$Cd to (60--75)\% were used
\cite{Rukhadze:2011b,Rukhadze:2011a,Rukhadze:2006}; now 23.2 g of
cadmium sample enriched in $^{106}$Cd to 99.57\% are installed in
the set-up \cite{Rukhadze:2016a}. The experiment gives the
strongest limit on the $2\nu 2$EC decay:
$T_{1/2}>4.7\times10^{20}$ yr. For other decay modes and channels
the sensitivity is at level of $10^{20}$ yr
\cite{Rukhadze:2016b}.

\begin{figure}
\centering
 \includegraphics[width=10 cm]{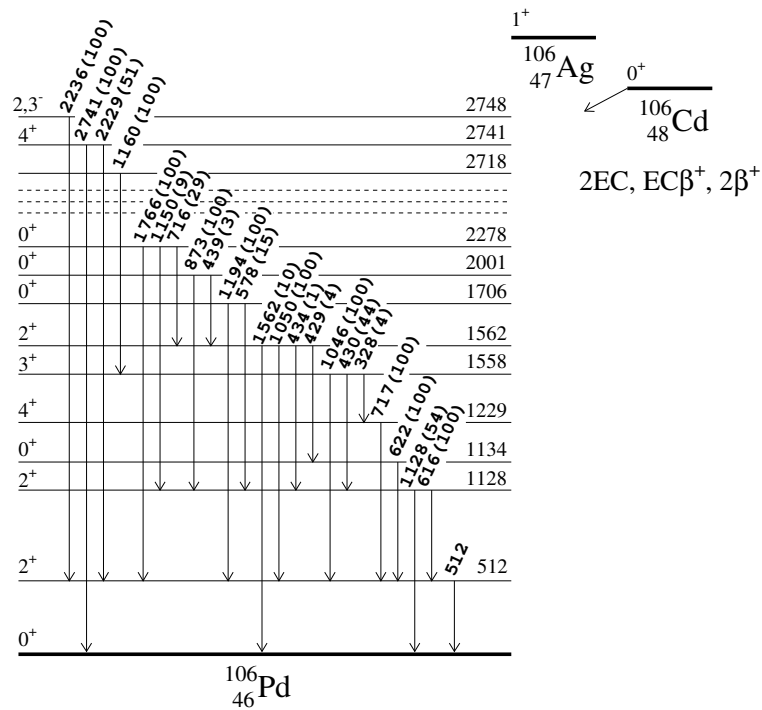}
\caption{Simplified decay scheme of $^{106}$Cd \cite{NDS106}
(levels with energies in the energy interval (2283--2714) keV are
omitted). Energies of the excited levels are in keV. Relative
intensities of $\gamma$ quanta are given in parentheses.}
 \label{fig:decay-scheme}
\end{figure}

A cadmium tungstate crystal scintillator from cadmium enriched in
$^{106}$Cd to 66\% ($^{106}$CdWO$_4$) was developed in 2010
\cite{Belli:2010}. The experiments with that detector are carried
out at the LNGS in the DAMA/CRYS, DAMA/R\&D set-ups, and in an
ultra-low background GeMulti HPGe $\gamma$ spectrometer of the
STELLA (SubTErranean Low Level Assay) facility
\cite{Laubenstein:2004} at the LNGS. The first stage of the
experiment with the $^{106}$CdWO$_4$ detector gave the half-life
limits on $2\beta$ processes in $^{106}$Cd at level of
$\sim10^{20}$ yr \cite{Belli:2012}. In the second stage the
$^{106}$CdWO$_4$ scintillator was installed between four HPGe
detectors (with volume $\simeq 225$ cm$^3$ each) of the GeMulti
HPGe $\gamma$ spectrometer to detect $\gamma$ quanta expected in
the most of the $^{106}$Cd decay channels, including the
annihilation $\gamma$'s emitted in decay modes with positron(s)
emission (a simplified decay scheme of $^{106}$Cd is presented in
Fig. \ref{fig:decay-scheme}). The experiment improved the
$^{106}$Cd half-life limits to the level of $T_{1/2} \geq
(10^{20}-10^{21})$ yr \cite{Belli:2016}. In the third stage,
described in the present report, the $^{106}$CdWO$_4$ detector was
running in coincidence (anti-coincidence) with two large volume
CdWO$_4$ crystal scintillators in a close geometry to increase the
detection efficiency to $\gamma$ quanta expected to be emitted
from the $^{106}$CdWO$_4$ crystal in the double beta decay
processes in $^{106}$Cd. Preliminary results of the experiment
stage were reported in \cite{Polischuk:2019}.

\section{The experiment}
\label{exp}

\begin{figure}
\centering
\includegraphics[width=14 cm]{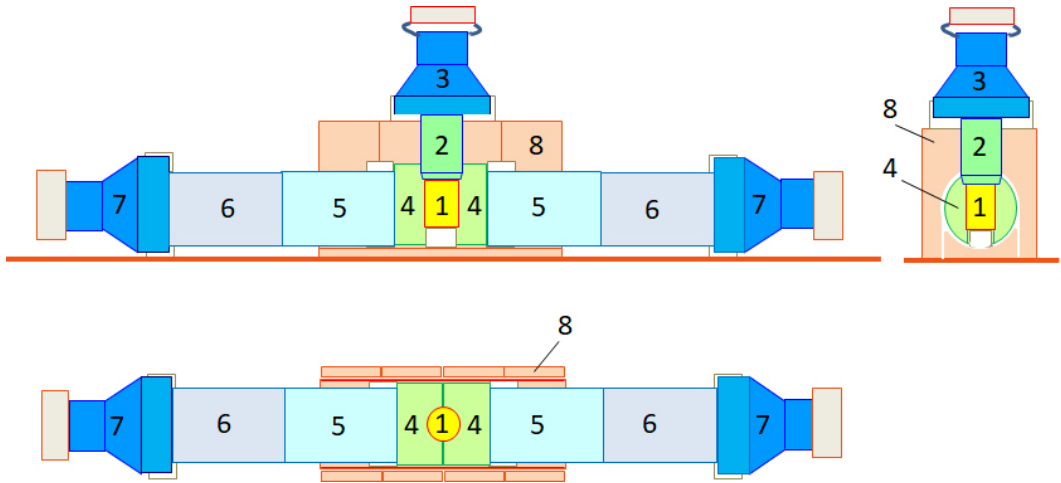}
\caption{Schematic of the experimental set-up with the
$^{106}$CdWO$_4$ scintillation detector. $^{106}$CdWO$_4$ crystal
scintillator (1) is viewed through PbWO$_4$ light-guide (2) by
photo-multiplier tube (3). Two CdWO$_4$ crystal scintillators (4)
are viewed through light-guides glued from quartz (5) and
polystyrene (6) by photo-multiplier tubes (7). The detector system
was surrounded by passive shield made from copper, lead,
polyethylene and cadmium (not shown). Only part of the copper
details (8, ``internal copper''), used to reduce the direct hits
of the detectors by $\gamma$ quanta from the PMTs, are shown. }
\label{fig:set-up}
\end{figure}

\begin{figure}
\resizebox{0.5\textwidth}{!}{\includegraphics{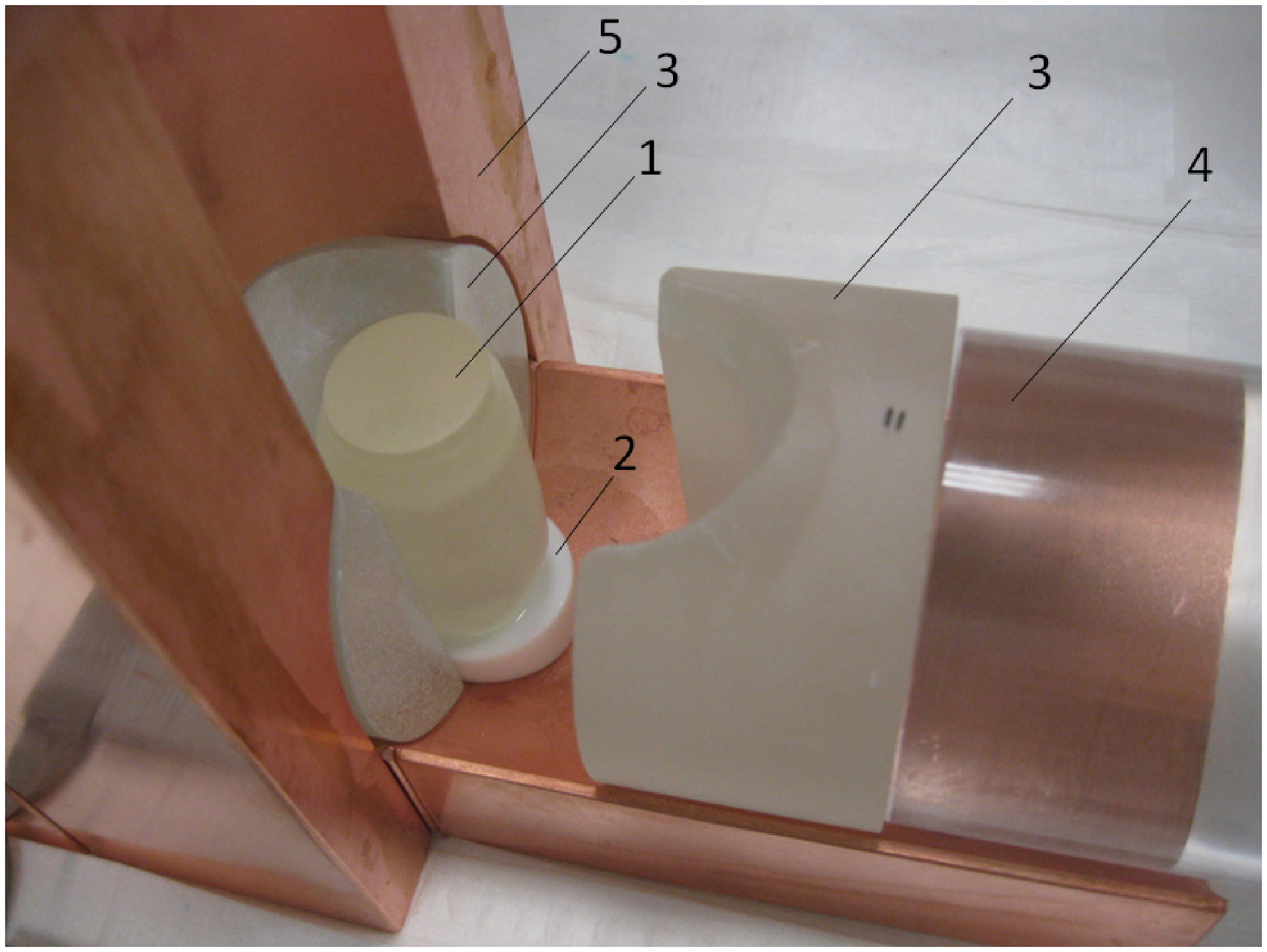}}
\resizebox{0.5\textwidth}{!}{\includegraphics{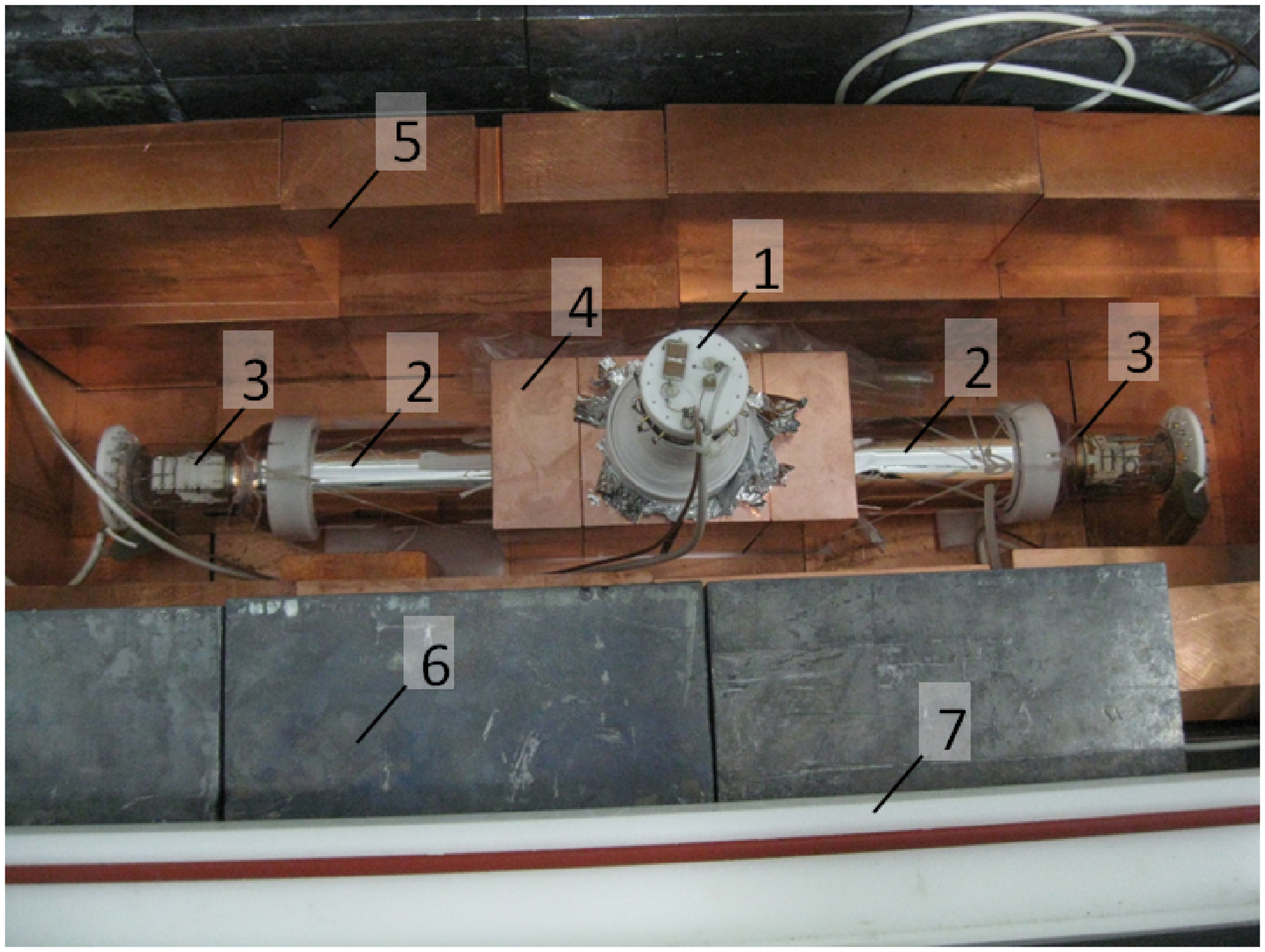}}
\caption{Left photograph: the $^{106}$CdWO$_4$
crystal scintillator (1), Teflon support of the $^{106}$CdWO$_4$
crystal (2), CdWO$_4$ crystal scintillators (3), quartz
light-guide (4), ``internal copper'' brick (5). Right photograph:
the detector system installed in the passive shield: PMT of the
$^{106}$CdWO$_4$ detector (1), light-guides of the CdWO$_4$
counters wrapped by reflecting foil (2), PMT of the CdWO$_4$ counters (3), ``internal
copper'' bricks (4), ``external copper'' bricks (5), lead bricks
(6), polyethylene shield (7). The copper, lead and polyethylene
shields are not completed.}
 \label{fig:photos}
\end{figure}

The $^{106}$CdWO$_4$ crystal scintillator of roughly cylindrical
shape (approximate sizes $\oslash 27$ mm $\times 50$ mm, mass
215.4 g) was viewed by a 3 inches low radioactive photo-multiplier
tube (PMT) Hamamatsu R6233MOD through a lead tungstate (PbWO$_4$)
crystal light-guide ($\oslash 40$ mm $\times 83$ mm). The PbWO$_4$
crystal has been developed from the highly purified
\cite{Boiko:2011} archaeological lead \cite{Danevich:2009}. Two
CdWO$_4$ crystal scintillators $\oslash 70$ mm $\times 38$ mm
include a cylindrical cut-out to house the $^{106}$CdWO$_4$
crystal. They were viewed by two 3 inches low radioactive PMTs
EMI9265B53/FL through light-guides glued in two parts: low
radioactive quartz ($\oslash 66$ mm $\times 100$ mm, close to the
CdWO$_4$ scintillators) and optical quality polystyrene ($\oslash
66$ mm $\times 100$ mm). A schematic of the set-up is shown in
Fig. \ref{fig:set-up}. The detector system was surrounded by four
high purity copper bricks (referred hereinafter as ``internal
copper'') and by layers of high purity copper (11 cm, referred
hereinafter as ``external copper''), low radioactive lead (10 cm),
cadmium (2 mm) and polyethylene (10 cm) to reduce the external
background. The inner volume of the set-up with the detector
system was continuously flushed by high-purity nitrogen gas to
remove environmental radon. The grade of the high-purity N$_2$ gas
is at least 5.5 for what concerns the presence of other possible
gases. However, the possible presence in trace of Radon gas in the
Nitrogen atmosphere inside the copper box, housing the detector,
has been checked with another set-up, by searching for the double
coincidences of the $\gamma$-rays (609 and 1120 keV) from
$^{214}$Bi Radon daughter. The obtained upper limit on the
possible Radon concentration in the high-purity Nitrogen
atmosphere has been measured to be: $<5.8\times10^{-2}$ Bq/m$^3$
(90\% C.L.) \cite{epjc2008}. Photographs of the detector system
are shown in Fig. \ref{fig:photos}.

An event-by-event data acquisition system based on a 100 MS/s 14
bit transient digitizer (DT5724 by CAEN) recorded the amplitude,
the arrival time and the pulse shape of each event. To reduce the
data volume due to presence in the $^{106}$CdWO$_4$ crystal of
$^{113}$Cd and $^{113m}$Cd $\beta$ active nuclides
\cite{Belli:2012,Belli:2010}, the energy threshold for the set-up
was set at level of $\approx510$ keV for the anti-coincidence
mode, while the energy threshold of the $^{106}$CdWO$_4$ detector
in the coincidence with the CdWO$_4$ counters was $\approx200$
keV. The energy thresholds of the CdWO$_4$ counters were
$\approx70$ keV. The energy scale and the energy resolution of the
detectors were measured with $^{22}$Na, $^{60}$Co, $^{133}$Ba,
$^{137}$Cs, and $^{228}$Th $\gamma$ sources at the beginning, in
the middle, and at the end of the experiment.

\begin{figure}
\centering
 \includegraphics[width=9 cm]{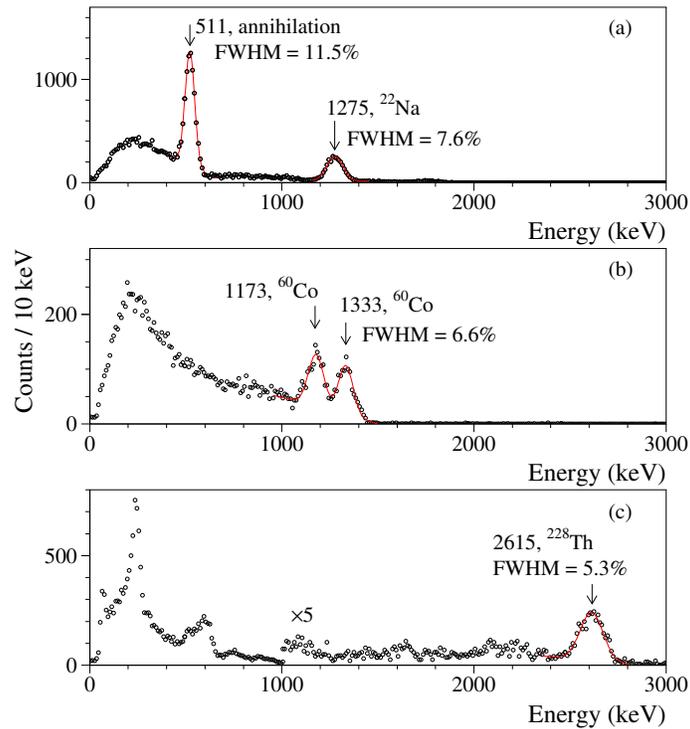}
\caption{Energy spectra of $^{22}$Na (a), $^{60}$Co (b) and
$^{228}$Th (c) $\gamma$ quanta measured by one of the CdWO$_4$
detectors. Fits of intensive $\gamma$ peaks by Gaussian functions
are shown by solid lines. Energies of $\gamma$ quanta are in keV.}
\label{fig:nat-CWO}
\end{figure}

The energy resolution of the $^{106}$CdWO$_4$ detector for the
total exposure can be described by the function $\mathrm{FWHM} =
6.85 \times \sqrt{E_{\gamma}}$, where FWHM (full width at half
maximum) and $E_{\gamma}$ are given in keV. The poor energy
resolution of the enriched detector (despite excellent optical
properties of the material \cite{Belli:2010}) is caused by the
elongated shape of the enriched scintillator that results in a
rather low and non-uniform light collection, and by using of not
perfectly transparent PbWO$_4$ crystal light-guide. The
performance of the CdWO$_4$ counters is substantially better.
The energy spectra accumulated by one of the counters with $^{22}$Na,
$^{60}$Co and $^{228}$Th $\gamma$ sources are presented in Fig.
\ref{fig:nat-CWO}. The energy resolution of the counters was
estimated by using the results of the three energy calibration
campaigns as $\mathrm{FWHM} = a \times \sqrt{E_{\gamma}} $ with
the coefficient $a$ equal to 2.97 and 3.13 for the two detectors. The resolution formulas take into account also energy scale shifts during the data taking over the experiment.

\begin{figure}
\centering
 \includegraphics[width=9 cm]{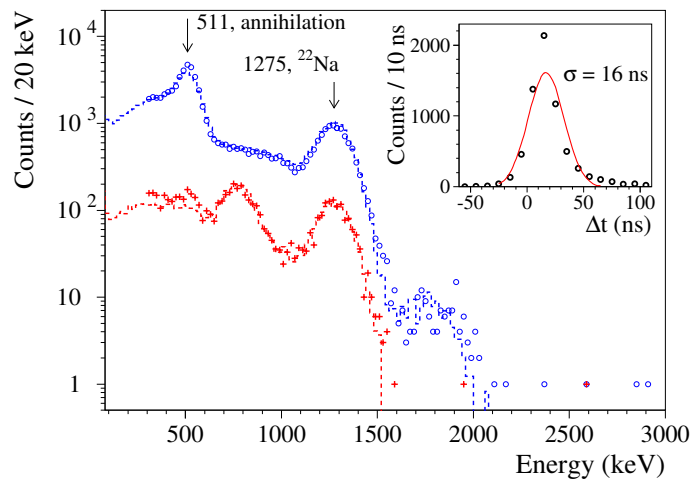}
\caption{Energy spectra of $^{22}$Na $\gamma$ quanta measured by
the $^{106}$CdWO$_4$ detector: with no coincidence cuts (blue
circles) and in coincidence with energy 511 keV in at least one of
the CdWO$_4$ counters (red crosses). The data simulated by using
the EGSnrc Monte Carlo code are drawn by dashed lines. (Inset)
Distribution of the $^{106}$CdWO$_4$ detector pulses start
positions relative to the CdWO$_4$ signals with the energy 511
keV.} \label{fig:sim-exp-22Na}
\end{figure}

Energy spectra of $^{22}$Na source were simulated by the EGSnrc
code \cite{EGS4}. The data measured with $^{22}$Na source without
coincidence selection and in coincidence with energy 511 keV in at
least one of the CdWO$_4$ counters is compared with the simulated
distribution in Fig. \ref{fig:sim-exp-22Na}. The experimental data
is in a reasonable agreement with the results of simulations.
% Some excess of the annihilation peak in the data can be explained by not perfectly implemented geometry of the experiment, particularly of the source holder, while annihilation process depends significantly on the presence of materials near the source.

A distribution of the $^{106}$CdWO$_4$ detector pulses start
positions relative to the CdWO$_4$ signals with energy 511 keV
is shown in Inset of Fig. \ref{fig:sim-exp-22Na}. The time
resolution of the detector system is rather high (the standard
deviation of the distribution is 16 ns) due to the fast rise time
of the CdWO$_4$ scintillation pulses.

\section{Results and Discussion}
\label{sec:res}

\subsection{Backgrounds reduction and model of the backgrounds}
\label{sec:BG-model}

The difference in CdWO$_4$ scintillation pulse shape for $\beta$ particles
($\gamma$ quanta) and $\alpha$ particles can be used to suppress the
background caused by $\alpha$ radioactive contamination of the
detector due to the residual contamination in $^{232}$Th and $^{238}$U with their daughters. The
mean time method was applied to the data to discriminate signals
of different origin by pulse shape. For each signal $f(t)$, the
numerical characteristic of its shape (mean time, $\zeta$) was
defined by using the following equation:

\begin{equation}
\zeta = \sum f(t_{k})\cdot t_{k} / \sum f(t_{k}),
\end{equation}

\noindent where the sum is over the time channels $k$, starting
from the origin of signal up to 35 $\mu$s; $f(t_{k})$ is the
digitized amplitude (at the time $t_{k}$) of a given signal. The
energy dependence of the parameter $\zeta$ and its standard
deviation (the distributions of $\zeta$ for $\beta$ particles
($\gamma$ quanta) and $\alpha$ particles are well described by a
Gaussian function) was determined by using the data of the
calibration measurements with $^{228}$Th gamma source. The
obtained parameters were then used to discriminate $\beta$
($\gamma$) events from $\alpha$ events in the data of the
low-background experiment. We refer reader to our previous works
\cite{Belli:2012,Belli:2016} where the pulse-shape discrimination
(PSD) method was described in detail.

\begin{figure}
\centering
 \includegraphics[width=11 cm]{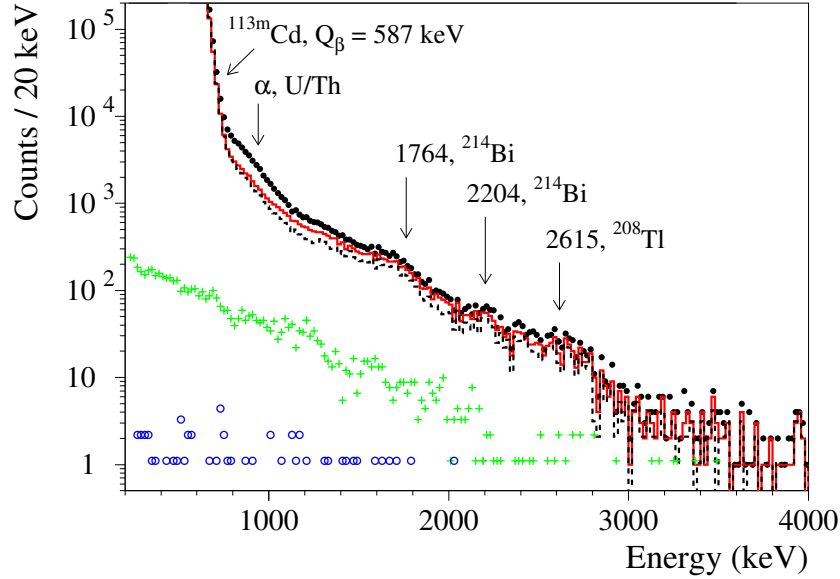}
\caption{Energy spectra measured by the $^{106}$CdWO$_4$ detector
for 26033 h in the low-background set-up without selection cuts
(black dots), after selection of $\gamma$ and $\beta$ events by
PSD using the mean time method (solid red line), the $\gamma$ and
$\beta$ events in anti-coincidence with the CdWO$_4$ counters
(dashed black line), the $\gamma$ and $\beta$ events in
coincidence with event(s) in at least one of the CdWO$_4$ counters
with the energy $E=511 \pm 2\sigma$ keV (green crosses), the
$\gamma$ and $\beta$ events in coincidence with events in both the
CdWO$_4$ counters with the energy $E=511 \pm 2\sigma$ keV (blue
circles).}
 \label{fig:diff-BG}
\end{figure}

By using the PSD the $\alpha$ events were statistically separated
from $\gamma(\beta)$ events. In addition the method discarded from
the data events of the $^{212}$Bi~--~$^{212}$Po sub-chain from the
$^{232}$Th family (due to the short decay time of $^{212}$Po
$\approx0.3$ $\mu$s these decays are treated by the data
acquisition system as a single event), PMT noise, pile-ups of
signals in the $^{106}$CdWO$_4$ detector, $^{106}$CdWO$_4$ plus
PbWO$_4$ events, etc. The results of the PSD method application to
the background data gathered for 26033 h in the low-background
set-up is shown in Fig. \ref{fig:diff-BG}. The mean time method
reduced the background mainly in the energy region (800--1300) keV
(by a factor $\sim1.6$) where $\alpha$ events of the $^{232}$Th
and $^{238}$U with their daughters are expected.

Further reduction of the background counting rate (by a factor
$\sim1.3$ in the energy interval (1000--3000) keV) was achieved
by exploiting the
anti-coincidence with the CdWO$_4$ counters. The background was
suppressed significantly by selection of events in the
$^{106}$CdWO$_4$ detector in coincidence with the event(s) in at
least one of the CdWO$_4$ counters with the energy release $E=511
\pm 2\sigma$ keV (by a factor $\sim17$ in the same energy interval; here $\sigma$ is the
energy resolution of the CdWO$_4$ counters for 511 keV $\gamma$
quanta), and by selection of events in coincidence with the events
in both the CdWO$_4$ counters with the energy $E=511 \pm 2\sigma$
keV (by a further factor $\sim42$). The stages of the background spectra
reduction are presented in Fig. \ref{fig:diff-BG}.

The counting rate of the $^{106}$CdWO$_4$ detector below the
energy of $\sim0.8$ MeV is mainly caused by the $\beta$ decay of
$^{113}$Cd with the energy $Q_{\beta}=323.83(27)$ keV
\cite{Wang:2017} and of $^{113m}$Cd ($Q_{\beta}=587.37(27)$ keV
\cite{Wang:2017,NDS113}). A background model to describe the
experimental data after the $^{113m}$Cd $\beta$ spectrum was
constructed from distributions of ``internal'' (radioactive
contamination of the $^{106}$CdWO$_4$ crystal) and ``external''
(radioactive contamination of the set-up details) sources. The
equilibrium of the $^{238}$U and $^{232}$Th chains in all the
materials is assumed to be broken\footnote{Secular equilibrium in
the $^{232}$Th and $^{238}$U decay families (when activities of
daughter nuclides are equal to the activity of their parent
nuclide) is typically broken in almost all materials due to
physical or chemical processes utilized in the material production
(see, e.g., \cite{Jagam:1993,Righi:2000,Danevich:2018}.}. The
sub-chains $^{228}$Ra$\rightarrow^{228}$Th,
$^{228}$Th$\rightarrow^{208}$Pb (the $^{232}$Th family) and
$^{238}$U$\rightarrow^{234}$U, $^{226}$Ra$\rightarrow^{210}$Pb,
$^{210}$Pb$\rightarrow^{206}$Pb ($^{238}$U) were assumed to be in
secular equilibrium.

The following ``internal'' sources were simulated in the
$^{106}$CdWO$_4$ crystal scintillator:

\begin{itemize}

 \item $^{40}$K, $^{228}$Ra$\rightarrow^{228}$Th,
    $^{228}$Th$\rightarrow^{208}$Pb,
    $^{226}$Ra$\rightarrow^{210}$Pb, and
    $^{210}$Pb$\rightarrow^{206}$Pb with activities estimated in the earlier stages of the experiment \cite{Danevich:2013,Poda:2013};
 \item distribution of $\alpha$ particles of $^{232}$Th and $^{238}$U with their daughters not discarded by the
 pulse-shape analysis;
 \item two-neutrino double beta decay of $^{116}$Cd with the
 half-life $T_{1/2}=2.63 \times 10^{19}$ yr \cite{Barabash:2018}.

\end{itemize}

The following ``external'' sources were simulated in the materials
of the set-up:

\begin{itemize}
    \item $^{40}$K, $^{228}$Ra$\rightarrow^{228}$Th,
    $^{228}$Th$\rightarrow^{208}$Pb,
    $^{226}$Ra$\rightarrow^{210}$Pb in the internal and external
    copper details, the quartz light guides, the PbWO$_4$ crystal
    light-guide, the PMTs;
    \item $^{210}$Pb$\rightarrow^{206}$Pb in the PbWO$_4$ crystal
    light-guide;
    \item $^{228}$Th$\rightarrow^{208}$Pb and $^{226}$Ra$\rightarrow^{210}$Pb in the CdWO$_4$ crystal
    scintillators;
    \item $^{56}$Co and $^{60}$Co in the internal copper bricks.
\end{itemize}

The background components were simulated using the EGSnrc package
with initial kinematics given by the DECAY0 event generator
\cite{Ponkratenko:2000}. The distribution of
residual $\alpha$ particles of $^{232}$Th and $^{238}$U with their
daughters was constructed from the experimental data by using the
pulse-shape analysis.

\begin{figure}
\centering
\includegraphics[width=10 cm]{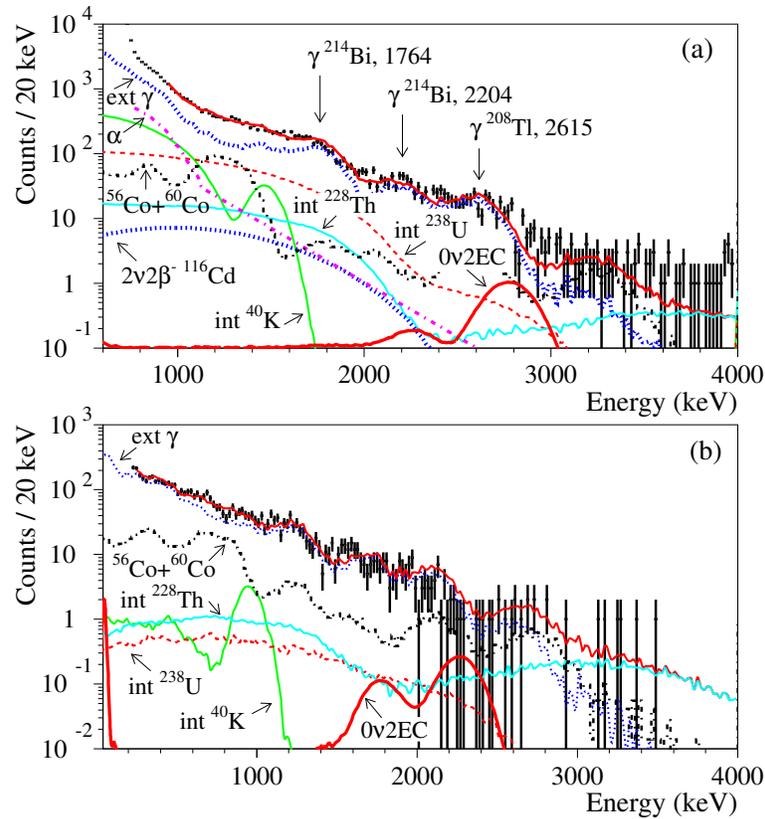}
\caption{Energy spectra of the $\gamma$ and $\beta$ events
accumulated for 26033 h by the $^{106}$CdWO$_4$ scintillation
detector in anti-coincidence with the CdWO$_4$ counters (a) and in
coincidence with the 511 keV annihilation $\gamma$ quanta in at least
one of the CdWO$_4$ counters (b) (points) together with the
background model (red line). The main components of the background
are shown: the distributions of internal contaminations (''int
$^{40}$K'', ''int $^{232}$Th'', and ''int $^{238}$U'') and
external $\gamma$ quanta (``ext $\gamma$''), residual $\alpha$
particles in the $^{106}$CdWO$_4$ crystal ($\alpha$), cosmogenic
$^{56}$Co and $^{60}$Co in the copper shield details, the
$2\nu2\beta$ decay of $^{116}$Cd. The excluded distributions of
the $0\nu2$EC decay of $^{106}$Cd to the ground state of
$^{106}$Pd with the half-life $T_{1/2}=6.8\times10^{20}$ yr are
shown by red solid line.}
 \label{fig:fit}
\end{figure}

The simulated models were used to fit the energy spectra of
$\gamma$ and $\beta$ events in anti-coincidence with the CdWO$_4$
counters and in coincidence with event(s) in at least one of the
CdWO$_4$ counters with the energy release $E=511 \pm 2\sigma$ keV.
The data were fitted in the energy intervals (940--4000) keV
(anti-coincidence data) and (240-3940) keV (coincidence with 511
keV). The fit quality is reasonable ($\chi^2=457$ for 235 degrees
of freedom). The results of the fit and the main components of the
background are shown in Fig. \ref{fig:fit}.

The fit allowed to estimate limits on radioactive contamination of
the materials of the low-background set-up. The data are presented
in Table \ref{tab:rad-cont}.

\begin{table}
\caption{Radioactive contamination (mBq/kg) of the materials of
the low-background set-up estimated by using the fit of the energy
spectra presented in Fig. \ref{fig:fit}. Upper limits are given at
68\% C.L.}
 \centering
\begin{tabular}{|l|l|l|l|l|l|l|l|l|}
 \hline
 Material           & $^{40}$K  & $^{56}$Co & $^{60}$Co & $^{88}$Y  & $^{210}$Pb            & $^{226}$Ra    & $^{228}$Ac    & $^{228}$Th \\
 \hline
 PbWO$_4$ crystal   & $\leq0.09$& --        & --        & --        & $\leq12\times10^3$    & $\leq0.07$    & $\leq0.28$    & $\leq0.23$ \\
 \hline
 CdWO$_4$ crystals  &  --       & --        & --        & --        & --                    & $\leq0.27$    & --            & $\leq0.014$ \\
 \hline
 Quartz light-guides& $\leq18$  & --        & --        & --        & --                    & $\leq3.3$     & $\leq0.6$     & $\leq0.6$ \\
 \hline
 Copper internal    & $\leq0.8$ & $\leq0.26$&$\leq0.5$  &$\leq0.005$& --                    & $\leq3.0$     & $\leq1.3$     & $\leq0.019$ \\
 \hline
 Copper external    & $\leq1.4$ & --        & --        & --        & --                    & $\leq1.5$     & $\leq3.2$     & $\leq0.026$ \\
 \hline
 PMTs               & $\leq1060$ & --        & --        & --        & --                   & $\leq140$     & $\leq1030$     & $\leq250$ \\
 \hline

\end{tabular}
 \label{tab:rad-cont}
\end{table}

\subsection{Limits on 2EC, EC$\beta^+$ and $2\beta^+$ processes in $^{106}$Cd}

There are no peculiarities in the experimental data that could be
ascribed to $2\beta$ processes in $^{106}$Cd. Lower limits on the
half-life of $^{106}$Cd relatively to different $2\beta$ decay
channels and modes can be estimated by using the following
formula:

\begin{equation}
\lim T_{1/2} = N \cdot \ln 2 \cdot \eta_\mathrm{{det}} \cdot
\eta_\mathrm{{sel}} \cdot t / \lim S,
\end{equation}

\noindent where $N$ is the number of $^{106}$Cd nuclei in the
$^{106}$CdWO$_4$ crystal ($2.42\times10^{23}$),
$\eta_\mathrm{{det}}$ is the detection efficiency for the process
of decay (calculated as a ratio of the events number in a
simulated distribution to the number of generated events),
$\eta_\mathrm{{sel}}$ is the selection cuts efficiency (selection
by PSD, time coincidence, energy interval), $t$ is the time of
measurements, and $\lim S$ is the number of events of the effect
searched for, which can be excluded at a given confidence level
(C.L.). The responses of the detector system to different modes and
channels of $^{106}$Cd double beta decay were simulated using the
EGSnrc package with initial kinematics given by the DECAY0 event
generator. About $5\times10^6$ events were generated for each
decay channel.

Different data were analyzed to estimate limits on the $2\beta$
processes in $^{106}$Cd. Fit of the anti-coincidence spectrum by
the above described model plus a simulated distribution of the
$0\nu2$EC decay of $^{106}$Cd to the ground state of $^{106}$Pd
returns the area of the distribution $(205\pm99)$ counts that is
no evidence for the effect searched for. According to
\cite{Feldman:1998} we took 367 events as $\lim S$ at 90\%
C.L.\footnote{In the present work all the limits are given with
90\% C.L. Only statistical errors coming from the data
fluctuations were taken into account in the estimations of the
$\lim S$ values, and systematic contributions have not been
included in the half-life limit values.} The detection efficiency
for the decay was simulated as $\eta_\mathrm{{det}}=0.522$. Taking
into account the selection cut efficiency due to application of
the PSD to select $\gamma$ and $\beta$ events
$\eta_\mathrm{{sel}}=0.955$, we got a lower limit on the half-life
of $^{106}$Cd relative to the $0\nu2$EC decay to the ground state
of $^{106}$Pd $T_{1/2}\geq6.8\times10^{20}$ yr (the excluded
distribution of the $0\nu$2EC decay is shown in Fig.
\ref{fig:fit}). The limit is slightly worse than the one obtained
in the previous stage of the experiment
($T_{1/2}\geq1.0\times10^{21}$ yr \cite{Belli:2012}, see also
Table \ref{tab:results}).

\begin{figure}
\centering
\includegraphics[width=10 cm]{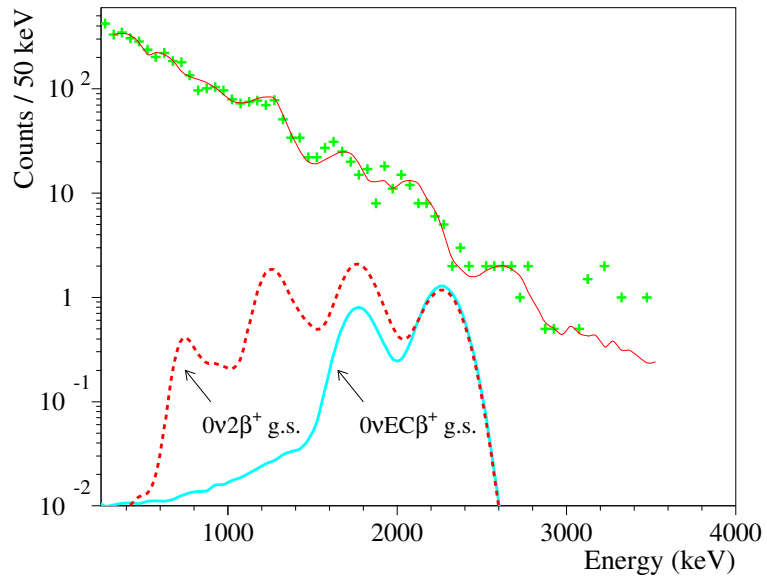}
\caption{Energy spectrum of the $\gamma$ and $\beta$ events
 measured for 26033 h by the $^{106}$CdWO$_4$ detector in coincidence with
events in at least one of the CdWO$_4$ counters with energy $E=511
\pm 2\sigma$ keV (crosses). The solid red line shows the fit of
the data by the background model (see Sec. \ref{sec:BG-model}).
Excluded distributions of $0\nu$EC$\beta^+$ and $0\nu2\beta^+$
decays of $^{106}$Cd to the ground state of $^{106}$Pd with the
half-lives $T_{1/2}=1.4\times10^{22}$ yr and
$T_{1/2}=5.9\times10^{21}$ yr, respectively, are shown.}
 \label{fig:CC511}
\end{figure}

Fit of the $^{106}$CdWO$_4$ detector data in coincidence with
signal(s) in the CdWO$_4$ counters by the above described
background model was more sensitive to the most of the modes and
channels of the decay searched for. An example of such an analysis
for the $0\nu$EC$\beta^+$ and $0\nu2\beta^+$ decays of $^{106}$Cd
to the ground state of $^{106}$Pd by using the data measured with
the $^{106}$CdWO$_4$ detector in coincidence with 511 keV events
in at least one of the CdWO$_4$ counters is shown in Fig.
\ref{fig:CC511}. The selection cuts efficiency, e.g., for the
$0\nu$EC$\beta^+$ process was calculated to be
$\eta_\mathrm{{sel}}=0.909$ as a product of the PSD to select
$\gamma$ and $\beta$ events in the interval $\pm 2\sigma$ of the
mean time values (0.9546), the time coincidence efficiency in the
interval $\pm 3\sigma$ (0.9973), the energy interval $\pm 2\sigma$
to select 511 keV events in the CdWO$_4$ counters (0.9545). The
data on the efficiencies, values of $\lim S$ and the obtained
half-life limits are given in Table \ref{tab:results}.

\begin{figure}
\centering
 \includegraphics[width=10 cm]{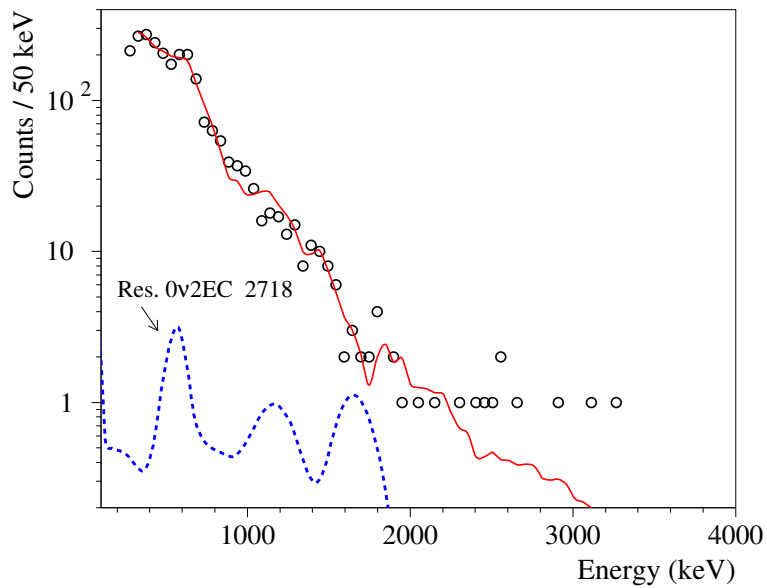}
\caption{Energy spectrum of $\gamma$ and $\beta$ events measured
by the $^{106}$CdWO$_4$ detector for 26033 h in coincidence with
event(s) in at least one of the CdWO$_4$ counters in the energy
interval $(1046-1.5\sigma$)--($1160 + 1.7\sigma$) keV (circles)
and its fit by the model of background (red line). The excluded
distribution of a possible resonant $0\nu2$EC decay of $^{106}$Cd to the 2718 keV
excited level of $^{106}$Pd with the half-life
$T_{1/2}=2.9\times10^{21}$ yr is shown.} \label{fig:CC1046-1160}
\end{figure}

Another example is search for $0\nu2$EC transition of $^{106}$Cd
to the 2718 keV excited level of $^{106}$Pd (considered as one of
the most promising decay channels from the point of view of a
possible resonant process \cite{Krivoruchenko:2011}). The search was
realized by analysis of the $^{106}$CdWO$_4$ detector data in
coincidence with event(s) in at least one of the CdWO$_4$ counters
in the energy interval $(1046-1.5\sigma$)--($1160 + 1.7\sigma$)
keV. The interval should contain two intensive $\gamma$ quanta
with energies 1046 keV and 1160 keV expected in the decay searched
for (see the decay scheme in Fig. \ref{fig:decay-scheme}).
The spectrum and its fit, consisting of the background model and excluded distribution
of the resonant process searched for, is presented in Fig. \ref{fig:CC1046-1160}.

\begin{figure}
\centering
 \includegraphics[width=10 cm]{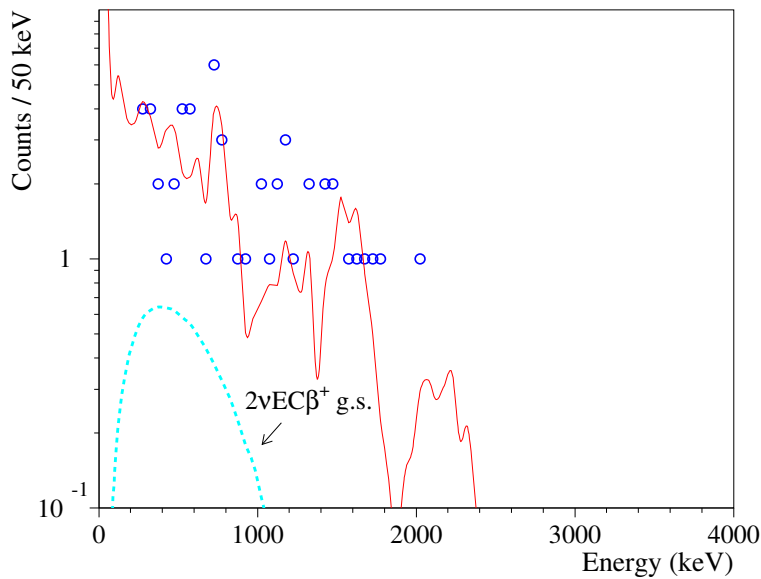}
\caption{Energy spectrum of $\gamma$ and $\beta$ events measured
by the $^{106}$CdWO$_4$ detector for 26033 h in coincidence with
511 keV annihilation $\gamma$ quanta in both of the CdWO$_4$
counters (circles). The expected background, built on the basis of
the fit presented in Fig. \ref{fig:fit}, is shown by red solid
line. The excluded distribution of the $2\nu$EC$\beta^+$ decay of
$^{106}$Cd to the ground state of $^{106}$Pd with the half-life
$T_{1/2}=2.1\times10^{21}$ yr is shown.}
 \label{fig:CC511-511}
\end{figure}

The highest sensitivity to several decay channels with positron(s)
emission was achieved by using the data gathered by the
$^{106}$CdWO$_4$ detector in coincidence with 511 keV annihilation
$\gamma$ quanta in both of the CdWO$_4$ counters thanks to a
rather high detection efficiency of the CdWO$_4$ counters and a
very low background counting rate (see Fig. \ref{fig:CC511-511}).
However, the fit of the spectrum by the background components is
not reliable enough due to a very low statistics of the data.
Thus, the method of comparison of the measured background with the
expected one was applied for the analysis. The expected background
was estimated from the results of the fit shown in Fig.
\ref{fig:fit}. There are 54 counts in the whole spectrum, while
the estimated background is 55.3 counts confirming a correct
background modelling. In the energy interval (250--1000) keV the
measured background is 33 counts, while the estimated one is 37.4
counts that leads to $\lim S=6.7$ counts in accordance with the
recommendations \cite{Feldman:1998}. Taking into account the
detection and the selections efficiencies for the
$2\nu$EC$\beta^+$ decay of $^{106}$Cd to the ground state of
$^{106}$Pd (0.040 and 0.703, respectively) one can get a
half-life limit $T_{1/2}=2.1\times10^{21}$ yr that is about two
times higher than the limit ($T_{1/2}=1.1\times10^{21}$ yr)
obtained in the previous stage of the experiment
\cite{Belli:2016}.

Limits on other $2\beta$ decay processes in $^{106}$Cd were
obtained in a similar way. They are presented in Table
\ref{tab:results}, where the results of the most sensitive
previous experiments are given for comparison.

\begin{table}
\caption{Half-life limits on $2\beta$ processes in $^{106}$Cd. The
experimental selection is also reported (AC, anti-coincidence; CC, in coincidence,
at the given energy (energies) with CdWO$_4$; CC 511\&511, in
coincidence with energies 511 keV in both of
the CdWO$_4$ counters). $\eta_\mathrm{{det}}$
denotes the detection efficiency, $\eta_\mathrm{{sel}}$ is the
selection cuts efficiency. The results of the most sensitive
previous experiments are given for comparison.}
 \centering
\begin{tabular}{|l|l|l|l|l|l|l|}
 \hline
  Decay,                        & Exp.              & $\eta_\mathrm{{det}}$  &  $\eta_\mathrm{{sel}}$ & $\lim S$  & \multicolumn{2}{c|}{ $\lim T_{1/2}$ (yr) at 90\% C.L.}\\
 \cline{6-7}
 level of $^{106}$Pd            & selection         &       &               &       & Present work              & Best previous \\
 \hline
 $2\nu2$EC 2$^+$ 1128           & CC 616            & 0.135 & 0.909        &  92   & $\geq6.6\times10^{20}$    & $\geq5.5\times10^{20}$ \cite{Belli:2016} \\
 \hline
 $2\nu2$EC 0$^+$ 1134           & CC 622            & 0.188 & 0.909        &  86   & $\geq9.9\times10^{20}$    & $\geq1.0\times10^{21}$ \cite{Belli:2016} \\
 \hline
 $2\nu2$EC 2$^+$ 1562           & CC 1050           & 0.138 & 0.909        &  80   & $\geq7.8\times10^{20}$    & $\geq7.4\times10^{20}$ \cite{Belli:2016} \\
 \hline
 $2\nu2$EC 0$^+$ 1706           & CC 1194           & 0.134 & 0.909        &  90   & $\geq6.7\times10^{20}$    & $\geq7.1\times10^{20}$ \cite{Belli:2016} \\
 \hline
 $2\nu2$EC 0$^+$ 2001           & CC 873            & 0.153 & 0.909        &  46   & $\geq1.5\times10^{21}$    & $\geq9.7\times10^{20}$ \cite{Belli:2016} \\
 \hline
 $2\nu2$EC 0$^+$ 2278           & CC 1766           & 0.091 & 0.909        &  131  & $\geq3.1\times10^{20}$    & $\geq1.0\times10^{21}$ \cite{Belli:2016} \\
 \hline
 $0\nu2$EC g.s.                 & AC                & 0.522 & 0.955        &  367  & $\geq6.8\times10^{20}$    & $\geq1.0\times10^{21}$ \cite{Belli:2012} \\
 \hline
 $0\nu2$EC 2$^+$ 512            & AC                & 0.319 & 0.955        &  443 & $\geq3.4\times10^{20}$    & $\geq5.1\times10^{20}$ \cite{Belli:2012} \\
 \hline
 $0\nu2$EC 2$^+$ 1128           & CC 616            & 0.118 & 0.909        &  110  & $\geq4.9\times10^{20}$    & $\geq5.1\times10^{20}$ \cite{Belli:2016} \\
 \hline
 $0\nu2$EC 0$^+$ 1134           & CC 622            & 0.155 & 0.909        &  109  & $\geq6.4\times10^{20}$    & $\geq1.1\times10^{21}$ \cite{Belli:2016} \\
 \hline
 $0\nu2$EC 2$^+$ 1562           & CC 1050           & 0.136 & 0.909        &  45   & $\geq1.4\times10^{21}$    & $\geq7.3\times10^{20}$ \cite{Belli:2016} \\
 \hline
 $0\nu2$EC 0$^+$ 1706           & CC 1194           & 0.120 & 0.909        &  27   & $\geq2.0\times10^{21}$    & $\geq1.0\times10^{21}$ \cite{Belli:2016} \\
 \hline
 $0\nu2$EC 0$^+$ 2001           & CC 873            & 0.135 & 0.909        &  177  & $\geq3.5\times10^{20}$    & $\geq1.2\times10^{21}$ \cite{Belli:2016} \\
 \hline
 $0\nu2$EC 0$^+$ 2278           & CC 1766           & 0.079 & 0.909        &  29   & $\geq1.2\times10^{21}$    & $\geq8.6\times10^{20}$ \cite{Belli:2016} \\
 \hline
 Res. $0\nu2$K 2718             & CC 1046 + 1160    & 0.215 & 0.909        &  33   & $\geq2.9\times10^{21}$    & $\geq1.1\times10^{21}$ \cite{Belli:2016} \\
 \hline
 Res. $0\nu KL_1$ 4$^+$ 2741    & AC                & 0.454 & 0.952         &  663  & $\geq3.2\times10^{20}$    & $\geq9.5\times10^{20}$ \cite{Belli:2012} \\
 \hline
 Res. $0\nu KL_3$ 2,3$^-$ 2748  & AC                & 0.318 & 0.955        &  432  & $\geq3.5\times10^{20}$    & $\geq1.4\times10^{21}$ \cite{Belli:2016} \\
 \hline
 $2\nu$EC$\beta^+$ g.s.         & CC 511\&511       & 0.040 & 0.703        &  6.7  & $\geq2.1\times10^{21}$    & $\geq1.1\times10^{21}$ \cite{Belli:2016} \\
 \hline
 $2\nu$EC$\beta^+$ 2$^+$ 512    & CC 511\&511       & 0.047 & 0.459        &  4.0  & $\geq2.7\times10^{21}$    & $\geq1.3\times10^{21}$ \cite{Belli:2016} \\
 \hline
 $2\nu$EC$\beta^+$ 2$^+$ 1128   & CC 511\&511       & 0.029 & 0.509        &  5.6  & $\geq1.3\times10^{21}$    & $\geq1.0\times10^{21}$ \cite{Belli:2016} \\
 \hline
 $2\nu$EC$\beta^+$ 0$^+$ 1134   & CC 511\&511       & 0.031 & 0.603        & 11    & $\geq8.5\times10^{20}$    & $\geq1.1\times10^{21}$ \cite{Belli:2016} \\
 \hline
 $0\nu$EC$\beta^+$ g.s.         & CC 511            & 0.376 & 0.909        & 12    & $\geq1.4\times10^{22}$    & $\geq2.2\times10^{21}$ \cite{Belli:2012} \\
 \hline
 $0\nu$EC$\beta^+$ 2$^+$ 512    & CC 511            & 0.384 & 0.909        & 18    & $\geq9.7\times10^{21}$    & $\geq1.9\times10^{21}$ \cite{Belli:2016} \\
 \hline
 $0\nu$EC$\beta^+$ 2$^+$ 1128   & CC 511            & 0.314 & 0.909        & 14    & $\geq1.0\times10^{22}$    & $\geq1.3\times10^{21}$ \cite{Belli:2016} \\
 \hline
 $0\nu$EC$\beta^+$ 0$^+$ 1134   & CC 511\&511       & 0.030 & 0.385        & 5.0   & $\geq1.2\times10^{21}$    & $\geq1.9\times10^{21}$ \cite{Belli:2016} \\
 \hline
 $2\nu2\beta^+$ g.s.            & CC 511\&511       & 0.052 & 0.385        &  5.8  & $\geq1.7\times10^{21}$    & $\geq2.3\times10^{21}$ \cite{Belli:2016} \\
 \hline
 $2\nu2\beta^+$  2$^+$ 512      & CC 511\&511       & 0.048 & 0.323        &  3.4  & $\geq2.3\times10^{21}$    & $\geq2.5\times10^{21}$ \cite{Belli:2016} \\
 \hline
 $0\nu2\beta^+$ g.s.            & CC 511            & 0.391 & 0.909        &  30   & $\geq5.9\times10^{21}$    & $\geq3.0\times10^{21}$ \cite{Belli:2016} \\
 \hline
 $0\nu2\beta^+$  2$^+$ 512      & CC 511            & 0.370 & 0.909        &  39   & $\geq4.0\times10^{21}$    & $\geq2.5\times10^{21}$ \cite{Belli:2016} \\
 \hline

\end{tabular}
 \label{tab:results}
\end{table}

A limit on effective nuclear matrix elements for the
$2\nu$EC$\beta^+$ decay of $^{106}$Cd to the ground state of
$^{106}$Pd could be estimated by using the calculations of the
phase-space factors for the decay \cite{Kotila:2013,Mirea:2015}
with the formula $1/T_{1/2} = G^{2\nu\mathrm{EC}\beta^+}\times
|M^{eff}|^2$. The effective matrix nuclear element $M^{eff}$ is
expressed by $M^{eff}= g^2_A \times M^{2\nu\mathrm{EC}\beta^+}$,
where $g_A$ is the axial-vector coupling constant,
$M^{2\nu\mathrm{EC}\beta^+}$ is nuclear matrix element. An upper
limit on the value of the effective matrix nuclear element for the
process can be estimated as $M^{eff}\leq(0.80-0.82)$.

The half-life limit on the $2\nu$EC$\beta^+$ decay of $^{106}$Cd
to the ground state of $^{106}$Pd, $T_{1/2} \geq 2.1\times10^{21}$
yr, approaches the region of the theoretical predictions that are
in the range $10^{21}-10^{22}$ yr
\cite{Hirsch:1994,Barabash:1996,Toivanen:1997,Rumyantsev:1998,Ejiri:2017}.
The sensitivity to the  double beta decay processes in $^{106}$Cd
is expected to be improved in the currently running experiment
with reduced background thanks to utilization of ultra-radiopure
PMTs, longer quartz light-guides for the CdWO$_4$ counters, a more
powerful passive shield of the detector system. Also the energy
resolution of the $^{106}$CdWO$_4$ detector was improved, roughly
by a factor $\sim1.8$, thanks to replacement of the PbWO$_4$
light-guide by a plastic scintillator light-guide with a
substantially better optical transmittance. This replacement
became possible due to an extremely low radioactive contamination
of the specially developed R11065-20 MOD Hamamatsu PMT
\cite{Bernabei:2012} used for the $^{106}$CdWO$_4$ detector.

\section{Conclusions}

The experiment to search for double beta decay of $^{106}$Cd with
enriched $^{106}$CdWO$_4$ scintillator in coincidence with two
large volume CdWO$_4$ scintillation counters was performed at the
Gran Sasso underground laboratory of INFN (Italy). New improved
limits are set on the different channels of $^{106}$Cd double beta
decay at level of $10^{20}-10^{22}$ yr. The new improved limit on
half-life of $^{106}$Cd relative to the $2\nu$EC$\beta^+$ decay
was estimated as $T_{1/2}\geq 2.1\times 10^{21}$ yr. The
sensitivity is within the region of the theoretical predictions
for the decay probability that are in the range of
$T_{1/2}\sim10^{21}-10^{22}$ yr. A new improved limit was set for
the resonant neutrinoless double-electron capture to the 2718 keV
excited level of $^{106}$Pd as
$T^{0\nu\mathrm{2K}}_{1/2}\geq2.9\times10^{21}$ yr.

The next stage of experiment is running at LNGS in the DAMA/R\&D
set-up with an improved sensitivity to all the decay channels
thanks to reduction of the background approximately by a factor
3--5 with utilization of ultra-radiopure PMTs, longer quartz
light-guides for the CdWO$_4$ counters, a more powerful passive
shield of the detector system. The energy resolution of the
$^{106}$CdWO$_4$ detector was improved too thanks to replacement
of the PbWO$_4$ light-guide by a plastic scintillator light-guide
with a substantially better optical transmittance. As a result,
the sensitivity to the $2\nu$EC$\beta^+$ decay of $^{106}$Cd is
expected to be high enough to detect the process with the
half-life at level of $\sim(0.5-1)\times10^{22}$ yr over 5 yr of
measurements.

\section{Acknowledgments}

D.V.K. and O.G.P. were supported in part by the project
``Investigation of double beta decay, rare alpha and beta decays''
of the program of the National Academy of Sciences of Ukraine
``Laboratory of young scientists'' Grant No. 0120U101838. F.A.D.,
D.V.K., V.R.K., V.V.K., V.I.T. and M.M.Z. were supported in part
by the project ``Double beta decay'' of the National Research
Foundation of Ukraine Grant No. 2020.02/0011. F.A.D. greatly
acknowledges the Government of Ukraine for the quarantine measures
that have been taken against the Coronavirus disease 2019 that
substantially reduced much unnecessary bureaucratic work.

\conflictsofinterest{The authors declare no conflict of interest.}

\vspace{0.8cm}
\noindent{\bf References}
\vspace{-1cm}

\externalbibliography{yes}
\bibliography{bibliography04}

\end{document}